# Early stage formation of graphene on the C-face of 6H-SiC

N. Camara,[1] G. Rius,[1] J.-R. Huntzinger,[2] A. Tiberj,[2] L. Magaud,[3] N. Mestres,[4]
P. Godignon,[1] and J. Camassel[2]

[1]*CNM-IMB-CSIC – Campus UAB 08193 Bellaterra, Barcelona, Spain*
[2]*GES – UMR 5650 Université Montpellier 2/CNRS, 34095 Montpellier cedex 5, France*
[3]*Institut Néel, CNRS and UJF, BP 166, 38042 Grenoble cedex 9, France*
[4]*ICMAB-CSIC, Campus UAB 08193 Bellaterra, Barcelona, Spain*

An investigation of the early stage formation of graphene on the C-face of 6H-silicon carbide (SiC) is presented. We show that the sublimation of few atomic layers of Si out of the SiC substrate is not homogeneous. In good agreement with the results of theoretical calculations it starts from defective sites, mainly dislocations that define nearly circular graphene layers, which have a pyramidal, volcano-like, shape with a center chimney where the original defect was located. At higher temperatures, complete conversion occurs but, again, it is not homogeneous. Within the sample surface, the intensity of the Raman bands, evidences non-homogeneous thickness. [DOI: ]

Graphene is a 2-dimensional carbon system with relativistic-like electronic transport properties [1] and the graphene-based devices should have the capability to breakthrough the silicon C-MOS leadership in the microelectronic industry. A widely used technique to fabricate mono or bi-layers of graphene was developed 4 years ago by exfoliating HOPG on an oxidized Si wafer. Unfortunately, the biggest flakes obtained in this way are only 10 x 10 $\mu m^2$, which is not enough for industrial purposes. An alternative technique is the use of chemical vapor deposition (CVD) deposition of carbon on a metal surface, like <0001> Ru or <111> Ir.[2,3] In this case the graphene transport properties are hidden by the metallic conductivity of the substrate and, for device application, complex transfer procedures are necessary. The last way of fabricating graphene is to sublimate few atomic layers of Si from a mono crystalline SiC substrate.[4,5] This can be done from a complete SiC wafer or from a small pre-patterned area,[6] and constitutes by far the most promising technique to develop industrial applications.[7] This is the one we present in this work.

Whatever the technique, the main concerns for graphene electronics are the properties of the substrate to graphene interface, the crystalline structure with emphasis on the long range order along the c-axis, the reproducibility and, finally, the homogeneity of layers. Two frequently used techniques to characterize the growth of few layers graphene (FLG) by SiC sublimation are low electron energy diffraction and scanning tunneling microscopy. To gain more realistic information on a wider area, optical microscopy (OM), atomic force microscopy (AFM), scanning electron microscopy (SEM) and Raman spectroscopy must be used.

In this work we investigate the formation of FLG grown on the C-face of a 6H-SiC substrate on which an atomically flat surface was prepared by Novasic.[8] Sublimation was done in a radio frequency induction furnace at temperatures ranging from 1450 to 1550°C. The processing time was about 5 min. Then, the growth product was systematically investigated by OM, SEM, AFM and Raman.

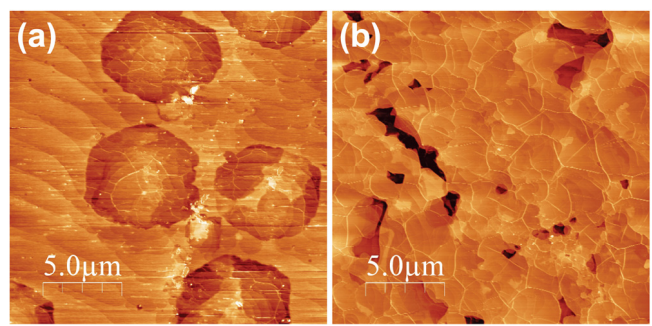

FIG. 1. Topographic AFM measurement in tapping mode of samples graphitized at (a) 1500°C, a typical 5μm diameter flake of FLG (confirmed by its Raman signature) is visible while outside the flake the SiC steps are visible, and (b) at 1550°C, the full wafer is covered by FLG (as determined by Raman spectroscopy).

We focus on three different types of samples obtained by different heating temperatures. They were 1x1 $cm^2$ templates cut from an on-axis 6H-SiC wafer, n-type doped to ~ $5x10^{17} cm^{-3}$. All chemical treatments used before sublimation was clean-room compatible and similar to the ones used before thermal oxidation. The vacuum limit was ~ $10^{-6}$ Torr. In order to remove any trace of native oxide, the temperature was raised to 1050°C for 10 minutes. Then, the three samples (A, B and C) were heated at 1450°C, 1500°C and 1550°C, respectively, for 5 minutes.

On the low temperature samples (Sample A) after heating at 1450°C for 5 minutes, we find that the only change is a large reconstruction of the initial surface. In the SiC literature, this surface reconstruction is best known as "step bunching" and originates from a small (non intentional) miscut of the nominally on-axis 6H-SiC wafer. It does not correspond to a single SiC bilayer (BL) height and, in the case of Sample A, resulted in large parallel terraces with ~ 1 nm height and 2 µm width. From OM, SEM or AFM no evidence of graphene is found. For the Raman spectroscopy measurements, we used the 514 nm line of an $Ar^+$ ion laser as exciting frequency and collected Raman spectra in the back-scattering configuration. The second order spectrum of the SiC substrate, with two main peaks at 1516 and 1714 cm$^{-1}$ at room temperature,[9] was used as internal reference. Despite intensive research, no graphene response from Raman investigations could be found. No evidence of Si aggregates before out-diffusion was neither found. This shows that working at 1450°C, under the pressure conditions used in this work, no efficient sublimation of Si atoms can occur. Upon heating, the surface only reorganizes to minimize energy.

On sample B, the situation is entirely different. After heating at 1500°C for 5 minutes, the first graphene layers appeared. They were not distributed homogeneously but rather randomly, with nearly circular shape (Fig. 1(a)) and a common diameter (~ 5 µm). The step-bunched SiC terraces already observed in sample A remained between the FLG areas. These layers were clearly evidenced by AFM (Fig. 1(a) and Fig. 2(c,d,e)), OM (Fig. 2(a)) and SEM (Fig 2(b)). The density over a full sample was about $10^6$ cm$^{-2}$, which is the typical density of dislocations in a commercial, research grade, SiC wafer. This suggests that the dislocations act as catalyzing defects, of which sublimated Si escapes more easily. With longer annealing times, the in-plane shape of the flakes turns from (nearly) circular to (roughly) hexagonal, with some pyramidal profile and a depression at the periphery.

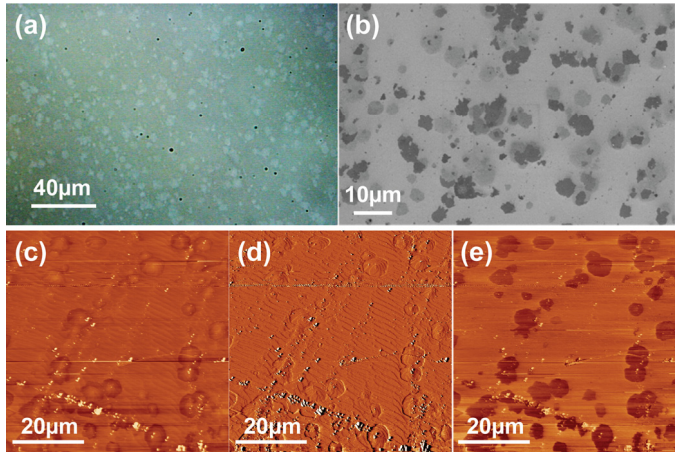

FIG. 2. Summary of results collected on sample B. (a) Wide range optical microscope view of the sample with the flakes clearly visible, (b) SEM picture of the same sample, the darker flakes visible at SEM are the brighter seen at the optical microscope and correspond to a higher number of graphene layers. (c), (d) and (e) are respectively the AFM measurement in tapping mode of the topography of the surface, the amplitude corresponding to contours, and the phase measurement.

If one considers that the growth starts from an extended defect, which acts like a "chimney" for a volcano, this is not so surprising. In this case, all constituting Si atoms in the topmost SiC BLs are progressively pumped by the central "chimney", until some frontier is reached. This can be just the diffusion length (but in this case the profile should remain flat at the periphery) or the boundary of the constitutive SiC grain. This is in better agreement with the final hexagonal shape and the depression at grain periphery, with no possible growth outside the periphery. This complex behavior was confirmed by Raman spectroscopy. Outside the flakes only the second-order Raman spectrum of the SiC substrate was detected while, inside the flakes we found the well-resolved graphite (G and 2D) bands, which give absolute evidence of carbon $sp^2$ re-organization.[10-12]

These observations suggest that there is a large energy barrier that prevents the (direct) diffusion of Si atoms through the topmost carbon layer. To cross-check this statement, we performed ab initio calculations to evaluate the energy difference for a Si atom located 1.99 Å below a graphene layer and moving in the layer. We assumed two different geometries: i) a perfect (infinite) honeycomb lattice; ii) a defective topmost layer with a Stone-Wales defect.[13] Both calculations were done using the code VASP[14] which is based on the density functional theory (DFT). We used the generalized gradient approximation and ultra-soft pseudopotentials that have been extensively tested.[15] Finally, a 3x3x3 k-point mesh was used so that, at convergence, the change of the total energy is below 0.001 eV. In the defect free carbon layer, the lateral (x, y) positions of the Si atom corresponded to the center of a honeycomb hexagon. With the Stone-Wales defect, it was fixed at the center of one of the defect heptagons. Both diffusion barriers calculated in this way were found very large. They are about 15.6 eV for the perfect layer and 9.7 eV in the presence of a Stone-Wales defect. The path through a defect is already lower but it is not enough to account for the strong difference shown in Fig. 1(a) and Fig. 1(b) for a 50°C temperature rise. This confirms that, in order to be evacuated, the direct jump of an in-depth Si atom to the topmost graphene layer is hardly possible. In other words, the fast sublimation of Si atoms, as observed in Fig. 1(a) and Fig. 2, requires a completely different process with a large assisting defect.

Finally, we investigated Sample C graphitized at 1550°C. The temperature rise was low but, again, the situation appears completely different. The whole surface is now covered by FLGs with no bare SiC. This appears clearly visible from the AFM picture shown in Fig. 1(b) or the OM picture shown in Fig. 3(a). Simply, the graphene thickness is still not homogeneous. The dark areas in Fig. 3(a) in crossed polarization mode microscopy are the "volcanos" already identified in Sample B. They are clearly associated to dislocations visible in dark field mode microscopy (Fig. 3(b)). The merged pictures of Fig. 3(a) and Fig. 3(b) revealed that the dislocations/chimneys are mainly responsible for the unhomogeneity in the graphene growth process (Fig. 3(c)). A clear continuum with less graphene and a rather smooth surface appears between the dark thicker FLG parts. This suggests a second (different) mechanism which, according to the results of the DFT calculation, should not be intrinsic but associated with a second type of defects. From the AFM measurements we find that all domain sizes are similar (in the range of few hundreds of nm) with, in some cases, incomplete coalescence (dark parts in Fig. 1(b)).

To confirm these results we used again Raman spectroscopy. From the ratio of the D and G peak intensity ($I_D/I_G$) we deduced the domain size of FLG using the empirical relation of Cancado et al.:[16]

$$L_a(nm) = \frac{560}{E_l^4}\left(\frac{I_D}{I_G}\right)^{-1} \quad (1)$$

Even in the worst case where the ratio $I_D/I_G$ is ~ 1/50 (see Fig. 3(d)) this gives crystallite size $L_a$ is in the range of 800 nm. This is in very reasonable agreement with the results of the AFM measurements.

Apart from the intensity which can change by a factor of 10 to 20 when moving from the continuum to the top of a chimney, the main difference from spectrum to spectrum comes from the 2D band. Being stacking order sensitive,[8,17] it should be used to follow the evolution from monolayer to bilayer or more complex turbostratic (with only one broad peak) to, finally, 3-dimensional HOPG (with two main bands).

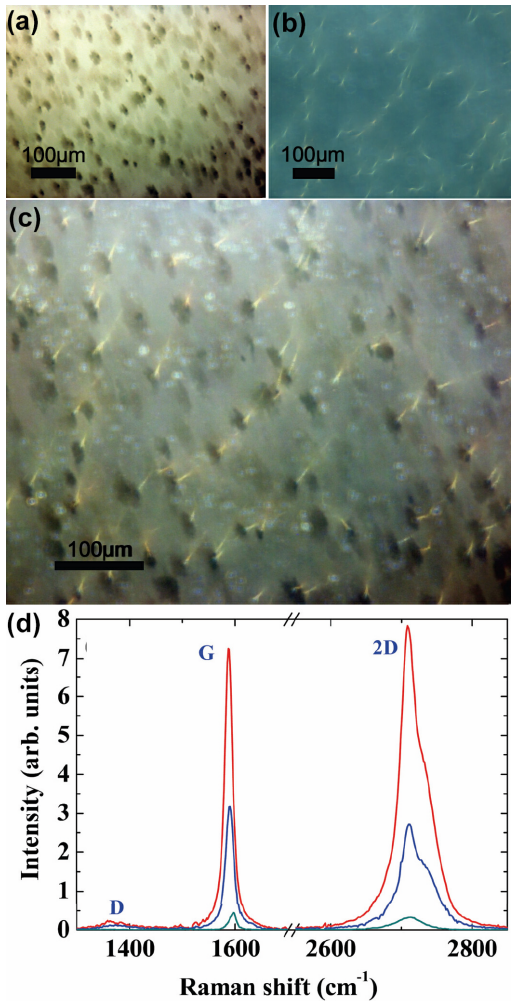

FIG. 3. (a) Wide range crossed polarized OM view of the sample C, graphitized at 1550°C and (b) the same area investigated in dark field mode. (c) is the superposition of (a) and (b) highlighting the corresponding between the thicker areas and the dislocations acting as chimneys. (d) Raman spectra collected on the same sample in three different parts: in the bright thin continuum (bottom spectrum), in the dark sublimation chimney (top spectrum) and in a grey intermediate part (middle). From bottom to top, the number of FLG increases by, typically, a factor of 10.

We found mainly two types of spectra. The first one appears only in the continuum (outside the flakes) giving evidence of thin and rather uniform FLG. The 2D band appears at 2710 cm$^{-1}$ and the full width at half maximum (FHWM) is ~ 42 cm$^{-1}$. The second series of spectra (inside the darker areas) not only confirms the larger thickness, it also indicates a different stacking. This may suggest that the speed of conversion plays a role in the 3-dimensional organization of the staging layers.

To summarize, in good agreement with the results of DFT calculations we have shown that the low pressure graphitization of 6H-SiC is not an intrinsic process. Within 50°C, on the C-face of 6H-SiC, two different mechanisms manifest. The first one involves the dislocations which are inherent to the limited quality of actual SiC substrates. The second gives more homogeneous results and more work is in progress to identify the participating defects. To grow large, homogeneous FLG, the first process will have to be eliminated.

We thank the French ANR and the European Community for partial support through the project. Financial support from the Spanish Government, NANOSELECT project and Juan de la Cierva grant to Dr. Nicolas Camara is also acknowledged.